\documentclass[onecolumn,11pt]{article}
\usepackage{supertabular,lscape,epsfig}
\usepackage{amssymb}
\usepackage{amsmath}
\usepackage{txfonts}
\usepackage{graphicx}
\usepackage{natbib}

\begin{document}

   \title{Interstellar medium in the M 43 nebula}

   \author{ {\bf Gnaci\'nski, Piotr }\\
   Institute of Theoretical Physics and Astrophysics,\\
              University of Gda\'nsk,
              ul. Wita Stwosza 57, 80-952 Gda\'nsk\\
              {\it pg@iftia9.univ.gda.pl }}
\maketitle

   \begin{abstract}
     We present a list of interstellar absorption lines in the direction of HD 37061 in the M 43 nebula. Some of the absorption lines arise from atomic excited levels that are uncommon in interstellar clouds. The excited levels of Fe II are populated by fluorescence. We found a large number of H$_2$ molecular absorption lines arising from vibrationally excited levels. The ortho/para H$_2$ ratio is equal to 2.7.
The H$_2$ rotational temperature of vibrational levels 1 -- 5 exceeds 2000 K.
   \end{abstract}
   {\bf  Key words: }{\it
   {ISM: clouds --- ISM: atoms --- ISM: lines --- ISM: molecules --- ultraviolet: ISM }
   }

\section{Introduction}
  
  The star HD 37061 is the ionization source for the H II region in the M 43 nebula.
M 43 itself is part of the Orion Nebula.
Beside the commonly observed interstellar lines, absorption lines from atomic excited states were observed in the Orion Nebula.
The sightlines to stars in the Orion Nebula show also anomalies in DIB strength (Jenniskens {\it et~al.}, 1994).
Some of the DIB strength anomalies are attributed to the presence of strong ionizing radiation.
 
  The sight line towards HD 37061 was one of the sight lines analyzed by Miller {\it et~al.} (2007). They have observed absorption from excited states of Fe II up to 1872.6 cm$^{-1}$.
Also the He I 3889 \AA\ absorption line from an excited level has been observed for many years.
In the direction of Trapezium the He I 3889 \AA\ line  has recently (Abel {\it et~al.}, 2006) been assigned to a separate ionisation layer localised 1.1 -- 1.6 pc from the
Trapezium.

\section{Interstellar absorption lines}

 \begin{figure}[tb]
    \centering
    \includegraphics[width=\textwidth,viewport=1 1 811 537,clip]{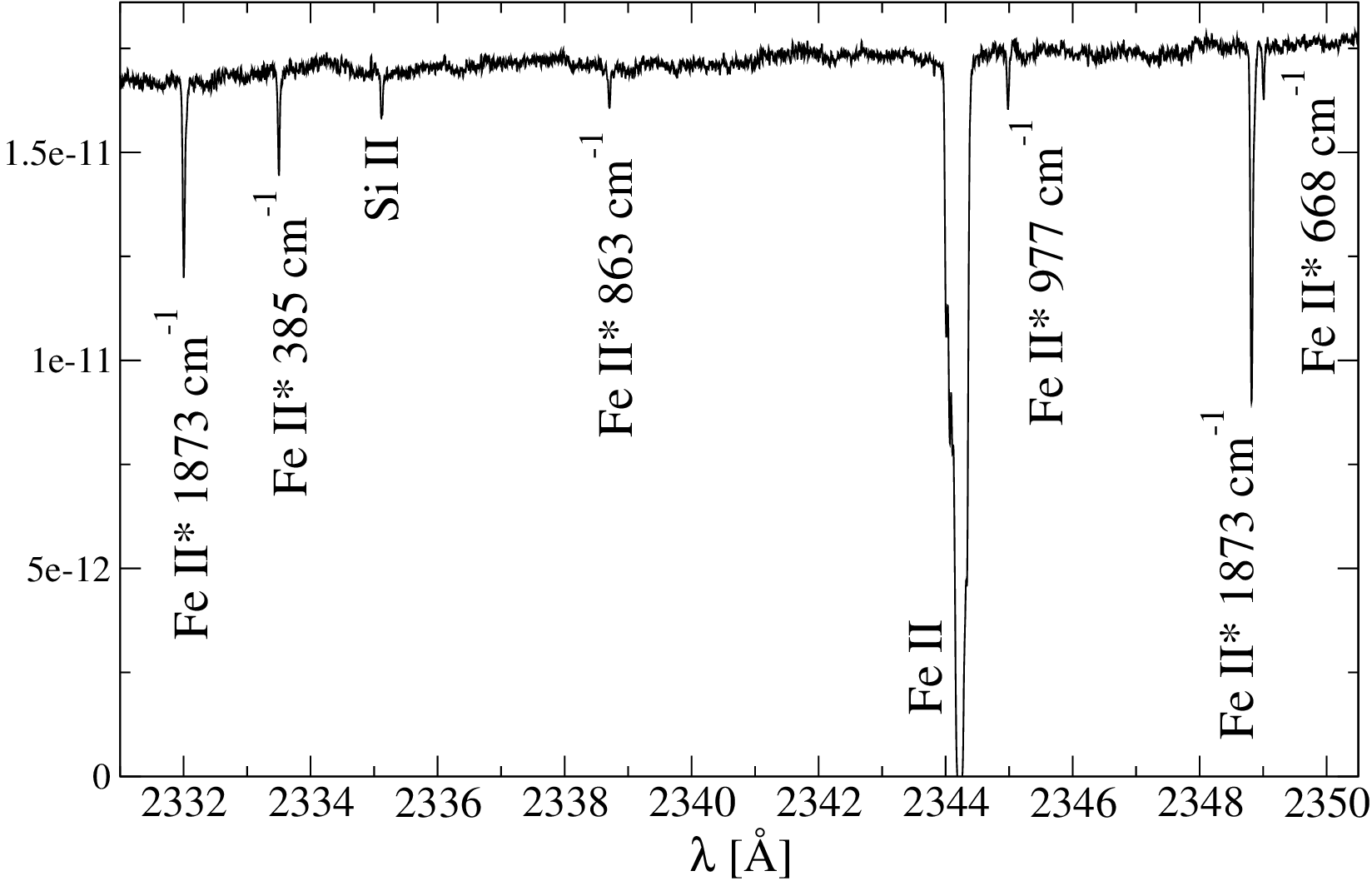}
    \caption{ 
      HST STIS spectrum of HD 37061 with iron absorption lines from excited states. Lower level energy (in cm$^{-1}$) is shown for excited states. 
    }
    \label{HD37061}
  \end{figure}
  
We have analyzed HST STIS (Space Telescope Imaging Spectrograph) spectra of HD 37061 covering the spectral range 1163 \AA\ -- 1356 \AA\ and 2128 \AA\ -- 2396 \AA, as well as visual UVES spectra covering the wavelength range 3050 \AA\ -- 10427 \AA. 

Table 1 presents a wavelength--ordered list of detected interstellar absorption lines in the
sightline of HD 37061. We list the electron configuration and energy of the lower level to
distinguish the absorption from the ground and excited electronic states.
The observed excited states are metastable with forbidden transitions to the
ground energy level.

The popular atomic data tables by Donald Morton (Morton (2000) and Morton (2003)) lists only lines from the ground states or an excited fine-structure state of the ground state.
All 8226 lines from the interstellar atomic lines compilation by Morton (2000) and Morton (2003) were considered as candidates for the interstellar features.
We have also considered 47681 atomic lines of H to Xe in neutral to doubly ionized states from the NIST ASD (National Institute of Standards and Technology -
Atomic Spectra Database, Ralchenko {\it et~al.}, 2008).

The interstellar Ca I 4227.9 \AA, CH$^+$ 4233.7 \AA\ and K I 7701 \AA\ absorption lines (vacuum wavelengths) usually observed in the visual spectral range are at the noise level.
The As II absorption line at 1263.77 \AA\ was first detected in a HST spectrum of $\zeta$ Ophiuchi by Cardelli {\it et~al.} (1993).

We have detected two molecules in the direction of HD 37061.
The H$_2$ absorption lines at 1274.922 \AA, 1276.325 \AA\ and 1335.131 \AA\ were observed. 
These lines were first reported by Federman {\it et~al.} (1995) in the spectrum of $\zeta$ Oph.
In total 39 lines of H$_2$ are clearly seen in the spectrum.
In the visual spectra range a CH line at 4301.5529 \AA\  was detected. The oscillator strength and wavelength for this line are adopted from Gredel {\it et~al.} (1993).
The CO and CH$^+$ lines were not detected.

Identifications of some absorption lines, like Cu II 2136.655 \AA\ and 2247.699 \AA\ are uncertain due to lack of oscillator strengths for those lines. The assignment of Cu II ion to these lines is based only on wavelength conformity.

{\footnotesize
  
\tablecaption{Interstellar absorption lines observed in the direction to HD 37061 (except H$_2$ lines).}
\tabletail{ \hline }
\tablelasttail{ \hline }
\tablefirsthead{
\hline 
 element  & wavelength [\AA]     & f (NIST)           & f (Morton) & E [cm$^{-1}$]        & configuration             & term    &   J   \\
 \hline 
}
\tablehead{
\hline 
 element  & wavelength [\AA]     & f (NIST)           & f (Morton) & E [cm$^{-1}$]        & configuration             & term    &   J   \\
 \hline 
}
\begin{supertabular}{lrllrlrc}
Mn II    &       1163.326 &               &        7.21e-3 &                0 & 3s$^2$ 3p$^6$ 3d$^5$ ($^6$S) 4s       & $^7$S      &   3   \\
Mn II    &       1164.208 &               &        4.70e-3 &                0 & 3s$^2$ 3p$^6$ 3d$^5$ ($^6$S) 4s       & $^7$S      &   3   \\
Ge II    &       1164.272 &               &        2.80e-1 &                0 & 4s$^2$ 4p                    & $^2$P$^o$     &  1/2  \\
Kr I     &       1164.867 &               &        1.88e-1 &                0 & 4s$^2$ 4p$^6$                   & $^1$S      &   0   \\
C I      &       1188.833 &        1.07e-2 &        1.24e-2 &                0 & 2s$^2$ 2p$^2$                   & $^3$P      &   0   \\
C I**    &       1189.447 &        3.36e-3 &        3.76e-3 &           43.414 & 2s$^2$ 2p$^2$                   & $^3$P      &   2   \\
C I**    &       1189.631 &        9.57e-3 &        1.12e-2 &           43.414 & 2s$^2$ 2p$^2$                   & $^3$P      &   2   \\
S III    &       1190.203 &        6.10e-1 &        2.37e-2 &                0 & 3s$^2$ 3p$^2$                   & $^3$P      &   0   \\
Si II    &       1190.416 &        2.77e-1 &        2.92e-1 &                0 & 3s$^2$ 3p                    & $^2$P$^o$     &  1/2  \\
C I*     &       1193.009 &        2.91e-2 &        2.81e-2 &           16.417 & 2s$^2$ 2p$^2$                   & $^3$P      &   1   \\
C I      &       1193.030 &        4.51e-2 &        4.09e-2 &                0 & 2s$^2$ 2p$^2$                   & $^3$P      &   0   \\
C I**    &       1193.240 &        3.65e-2 &        3.32e-2 &           43.414 & 2s$^2$ 2p$^2$                   & $^3$P      &   2   \\
C I*     &       1193.264 &        1.03e-2 &        9.15e-3 &           16.417 & 2s$^2$ 2p$^2$                   & $^3$P      &   1   \\
Si II    &       1193.290 &        5.75e-1 &        5.82e-1 &                0 & 3s$^2$ 3p                    & $^2$P$^o$     &  1/2  \\
C I*     &       1193.679 &        1.02e-2 &        9.05e-3 &           16.417 & 2s$^2$ 2p$^2$                   & $^3$P      &   1   \\
C I      &       1193.995 &        1.28e-2 &        1.24e-2 &                0 & 2s$^2$ 2p$^2$                   & $^3$P      &   0   \\
S III*   &       1194.058 &        4.60e-1 &        1.77e-2 &           298.69 & 3s$^2$ 3p$^2$                   & $^3$P      &   1   \\
C I*     &       1194.063 &        5.84e-3 &        6.63e-3 &           43.414 & 2s$^2$ 2p$^2$                   & $^3$P      &   2   \\
S III*   &       1194.449 &        1.50e-1 &        5.89e-3 &           298.69 & 3s$^2$ 3p$^2$                   & $^3$P      &   1   \\
Si II*   &       1194.500 &        7.37e-1 &        7.27e-1 &           287.24 & 3s$^2$ 3p                    & $^2$P$^o$     &  3/2  \\
Mn II    &       1197.184 &               &        2.17e-1 &                0 & 3s$^2$ 3p$^6$ 3d$^5$ ($^6$S) 4s       & $^7$S      &   3   \\
Si II*   &       1197.394 &        1.50e-1 &        1.45e-1 &           287.24 & 3s$^2$ 3p                    & $^2$P$^o$     &  3/2  \\
Mn II    &       1199.391 &               &        1.69e-1 &                0 & 3s$^2$ 3p$^6$ 3d$^5$ ($^6$S) 4s       & $^7$S      &   3   \\
N I      &       1199.550 &        1.30e-1 &        1.32e-1 &                0 & 2s$^2$ 2p$^3$                   & $^4$S$^o$     &  3/2  \\
N I      &       1200.223 &        8.62e-2 &        8.69e-2 &                0 & 2s$^2$ 2p$^3$                   & $^4$S$^o$     &  3/2  \\
N I      &       1200.710 &        4.30e-2 &        4.32e-2 &                0 & 2s$^2$ 2p$^3$                   & $^4$S$^o$     &  3/2  \\
S III*   &       1200.966 &        5.10e-1 &        1.97e-2 &           833.08 & 3s$^2$ 3p$^2$                   & $^3$P      &   2   \\
Mn II    &       1201.118 &               &        1.21e-1 &                0 & 3s$^2$ 3p$^6$ 3d$^5$ ($^6$S) 4s       & $^7$S      &   3   \\
S III*   &       1201.726 &        9.10e-2 &        3.51e-3 &           833.08 & 3s$^2$ 3p$^2$                   & $^3$P      &   2   \\
Si III   &       1206.500 &        1.67 &        1.63 &                0 & 3s$^2$                       & $^1$S      &   0   \\
Kr I     &       1235.838 &               &        2.04e-1 &                0 & 4s$^2$ 4p$^6$                   & $^1$S      &   0   \\
Ge II    &       1237.059 &        8.70e-1 &        1.23 &                0 & 4s$^2$ 4p                    & $^2$P$^o$     &  1/2  \\
Mg II    &       1239.925 &        6.21e-4 &        6.32e-4 &                0 & 2p$^6$ 3s                    & $^2$S      &  1/2  \\
Mg II    &       1240.395 &        3.51e-4 &        3.56e-4 &                0 & 2p$^6$ 3s                    & $^2$S      &  1/2  \\
N I**    &       1243.171 &        5.16e-3 &               &        19224.464 & 2s$^2$ 2p$^3$                   & $^2$D$^o$     &  5/2  \\
N I**    &       1243.179 &        7.44e-2 &               &        19224.464 & 2s$^2$ 2p$^3$                   & $^2$D$^o$     &  5/2  \\
N I*     &       1243.306 &        7.16e-2 &               &        19233.177 & 2s$^2$ 2p$^3$                   & $^2$D$^o$     &  3/2  \\
N I*     &       1243.313 &        8.17e-3 &               &        19233.177 & 2s$^2$ 2p$^3$                   & $^2$D$^o$     &  3/2  \\
S II     &       1250.578 &        5.40e-3 &        5.43e-3 &                0 & 3s$^2$ 3p$^3$                   & $^4$S$^o$     &  3/2  \\
S II     &       1253.805 &        9.90e-3 &        1.09e-2 &                0 & 3s$^2$ 3p$^3$                   & $^4$S$^o$     &  3/2  \\
S II     &       1259.518 &        1.20e-2 &        1.66e-2 &                0 & 3s$^2$ 3p$^3$                   & $^4$S$^o$     &  3/2  \\
Si II    &       1260.422 &        1.22 &        1.18 &                0 & 3s$^2$ 3p                    & $^2$P$^o$     &  1/2  \\
Fe II    &       1260.533 &               &        2.40e-2 &                0 & 3d$^6$ ($^5$D) 4s               & a $^6$D    &  9/2  \\
C I      &       1260.735 &        3.80e-2 &        5.07e-2 &                0 & 2s$^2$ 2p$^2$                   & $^3$P      &   0   \\
C I*     &       1260.926 &        1.35e-2 &        1.75e-2 &           16.417 & 2s$^2$ 2p$^2$                   & $^3$P      &   1   \\
C I*     &       1260.996 &        1.05e-2 &        1.34e-2 &           16.417 & 2s$^2$ 2p$^2$                   & $^3$P      &   1   \\
C I*     &       1261.122 &        1.47e-2 &        2.02e-2 &           16.417 & 2s$^2$ 2p$^2$                   & $^3$P      &   1   \\
C I**    &       1261.426 &        1.01e-2 &        1.31e-2 &           43.414 & 2s$^2$ 2p$^2$                   & $^3$P      &   2   \\
C I**    &       1261.552 &        3.03e-2 &        3.91e-2 &           43.414 & 2s$^2$ 2p$^2$                   & $^3$P      &   2   \\
As II    &       1263.770 &               &        2.59e-1 &                0 & 4s$^2$ 4p$^2$                   & $^3$P      &   0   \\
Si II*   &       1264.738 &        1.09 &        1.05 &           287.24 & 3s$^2$ 3p                    & $^2$P$^o$     &  3/2  \\
Si II*   &       1265.002 &        1.13e-1 &        1.17e-1 &           287.24 & 3s$^2$ 3p                    & $^2$P$^o$     &  3/2  \\
C I      &       1277.245 &        9.32e-2 &        8.53e-2 &                0 & 2s$^2$ 2p$^2$                   & $^3$P      &   0   \\
C I*     &       1277.283 &        7.05e-2 &        6.66e-2 &           16.417 & 2s$^2$ 2p$^2$                   & $^3$P      &   1   \\
C I*     &       1277.513 &        2.23e-2 &        2.10e-2 &           16.417 & 2s$^2$ 2p$^2$                   & $^3$P      &   1   \\
C I**    &       1277.550 &        7.91e-2 &        7.63e-2 &           43.414 & 2s$^2$ 2p$^2$                   & $^3$P      &   2   \\
C I**    &       1277.723 &        1.55e-2 &        1.53e-2 &           43.414 & 2s$^2$ 2p$^2$                   & $^3$P      &   2   \\
C I*     &       1279.056 &        7.36e-4 &        7.08e-4 &           16.417 & 2s$^2$ 2p$^2$                   & $^3$P      &   1   \\
C I**    &       1279.229 &        3.78e-3 &        2.14e-3 &           43.414 & 2s$^2$ 2p$^2$                   & $^3$P      &   2   \\
C I*     &       1279.891 &        1.26e-2 &        1.43e-2 &           16.417 & 2s$^2$ 2p$^2$                   & $^3$P      &   1   \\
C I      &       1280.135 &        2.29e-2 &        2.63e-2 &                0 & 2s$^2$ 2p$^2$                   & $^3$P      &   0   \\
C I**    &       1280.333 &        1.42e-2 &        1.52e-2 &           43.414 & 2s$^2$ 2p$^2$                   & $^3$P      &   2   \\
C I*     &       1280.404 &        4.25e-3 &        4.40e-3 &           16.417 & 2s$^2$ 2p$^2$                   & $^3$P      &   1   \\
C I*     &       1280.598 &        6.74e-3 &        7.04e-3 &           16.417 & 2s$^2$ 2p$^2$                   & $^3$P      &   1   \\
C I**    &       1280.847 &        4.91e-3 &        5.22e-3 &           43.414 & 2s$^2$ 2p$^2$                   & $^3$P      &   2   \\
S I      &       1295.653 &        1.20e-1 &        8.70e-2 &                0 & 3s$^2$ 3p$^4$                   & $^3$P      &   2   \\
P II     &       1301.874 &        3.80e-2 &        1.27e-2 &                0 & 3s$^2$ 3p$^2$                   & $^3$P      &   0   \\
O I      &       1302.169 &        5.20e-2 &        4.80e-2 &                0 & 2s$^2$ 2p$^4$                   & $^3$P      &   2   \\
Si II    &       1304.370 &        9.28e-2 &        8.63e-2 &                0 & 3s$^2$ ($^1$S) 3p               & $^2$P$^o$     &  1/2  \\
O I*     &       1304.858 &        5.18e-2 &        4.78e-2 &          158.265 & 2s$^2$ 2p$^4$                   & $^3$P      &   1   \\
O I**    &       1306.029 &        5.19e-2 &        4.78e-2 &          226.977 & 2s$^2$ 2p$^4$                   & $^3$P      &   0   \\
Ni II    &       1308.866 &               &               &                0 & 3s$^2$ 3p$^6$ 3d$^9$               & $^2$D      &  5/2  \\
Si II*   &       1309.276 &        8.00e-2 &        8.60e-2 &           287.24 & 3s$^2$ 3p                    & $^2$P$^o$     &  3/2  \\
Ni II    &       1317.217 &               &               &                0 & 3s$^2$ 3p$^6$ 3d$^9$               & $^2$D      &  5/2  \\
C I      &       1328.833 &        6.31e-2 &        7.58e-2 &                0 & 2s$^2$ 2p$^2$                   & $^3$P      &   0   \\
C I*     &       1329.085 &        2.13e-2 &        2.54e-2 &           16.417 & 2s$^2$ 2p$^2$                   & $^3$P      &   1   \\
C I*     &       1329.100 &        2.60e-2 &        3.13e-2 &           16.417 & 2s$^2$ 2p$^2$                   & $^3$P      &   1   \\
C I*     &       1329.123 &        1.60e-2 &        1.91e-2 &           16.417 & 2s$^2$ 2p$^2$                   & $^3$P      &   1   \\
C I**    &       1329.578 &        4.74e-2 &        5.69e-2 &           43.414 & 2s$^2$ 2p$^2$                   & $^3$P      &   2   \\
C I**    &       1329.600 &        1.59e-2 &        1.89e-2 &           43.414 & 2s$^2$ 2p$^2$                   & $^3$P      &   2   \\
C II     &       1334.519 &        1.27e-1 &        1.28e-1 &                0 & 2s$^2$ ($^1$S) 2p               & $^2$P$^o$     &  1/2  \\
P III    &       1334.813 &        2.90e-2 &        2.82e-2 &                0 & 3s$^2$ ($^1$S) 3p               & $^2$P$^o$     &  1/2  \\
C II*    &       1335.649 &        1.27e-2 &        1.28e-2 &            63.42 & 2s$^2$ ($^1$S) 2p               & $^2$P$^o$     &  3/2  \\
C II*    &       1335.692 &        1.14e-1 &        1.15e-1 &            63.42 & 2s$^2$ ($^1$S) 2p               & $^2$P$^o$     &  3/2  \\
Ni II    &       1345.878 &               &        7.69e-3 &                0 & 3s$^2$ 3p$^6$ 3d$^9$               & $^2$D      &  5/2  \\
Cl I     &       1347.240 &        1.14e-1 &        1.53e-1 &                0 & 3s$^2$ 3p$^5$                   & $^2$P$^o$     &  3/2  \\
O I      &       1355.598 &        1.16e-6 &        1.16e-6 &                0 & 2s$^2$ 2p$^4$                   & $^3$P      &   2   \\
Cu II    &       2136.655 &               &               &                  &                           &         &       \\
Cd II    &       2145.070 &        3.90e-1 &        4.98e-1 &                0 & 5s                        & $^2$S      &  1/2  \\
Ni II*   &       2166.230 &        1.70e-1 &               &           8393.9 & 3p$^6$ 3d$^8$ ($^3$F) 4s           & $^4$F      &  9/2  \\
Fe I     &       2167.453 &        1.50e-1 &        1.50e-1 &                0 & 3p$^6$ 3d$^6$ 4s$^2$               & a $^5$D    &   4   \\
Ni II*   &       2217.167 &        3.00e-1 &               &           8393.9 & 3p$^6$ 3d$^8$ ($^3$F) 4s           & $^4$F      &  9/2  \\
Ni II*   &       2223.642 &        7.30e-2 &               &           8393.9 & 3p$^6$ 3d$^8$ ($^3$F) 4s           & $^4$F      &  9/2  \\
Fe II    &       2234.447 &               &        2.52e-5 &                0 & 3d$^6$ ($^5$D) 4s               & a $^6$D    &  9/2  \\
Cu II    &       2247.699 &               &               &                  &                           &         &       \\
Fe II    &       2249.877 &        2.50e-3 &        1.82e-3 &                0 & 3d$^6$ ($^5$D) 4s               & a $^6$D    &  9/2  \\
Fe II    &       2260.781 &        3.80e-3 &        2.44e-3 &                0 & 3d$^6$ ($^5$D) 4s               & a $^6$D    &  9/2  \\
Cd II    &       2265.715 &        2.30e-1 &        2.47e-1 &                0 & 5s                        & $^2$S      &  1/2  \\
Co II*   &       2286.856 &        3.10e-1 &               &          3350.58 & 3p$^6$ 3d$^7$ (4F) 4s           & a $^5$F    &   5   \\
Mn II    &       2299.663 &               &        4.81e-4 &                0 & 3s$^2$ 3p$^6$ 3d$^5$ ($^6$S) 4s       & $^7$S      &   3   \\
Mn II    &       2305.714 &               &        1.15e-3 &                0 & 3s$^2$ 3p$^6$ 3d$^5$ ($^6$S) 4s       & $^7$S      &   3   \\
Ni II*   &       2316.748 &        1.85e-1 &               &           8393.9 & 3p$^6$ 3d$^8$ ($^3$F) 4s           & $^4$F      &  9/2  \\
C II     &       2325.403 &        5.92e-8 &        4.78e-8 &                0 & 2s$^2$ ($^1$S) 2p               & $^2$P$^o$     &  1/2  \\
Fe II*   &       2332.022 &        1.90e-2 &               &         1872.567 & 3p$^6$ 3d$^7$                   & a $^4$F    &  9/2  \\
Fe II*   &       2333.516 &        9.20e-2 &        7.78e-2 &           384.79 & 3d$^6$ ($^5$D) 4s               & a $^6$D    &  7/2  \\
Si II    &       2335.123 &        4.50e-6 &        4.25e-6 &                0 & 3s$^2$ ($^1$S) 3p               & $^2$P$^o$     &  1/2  \\
Fe II*   &       2338.725 &        9.00e-2 &        8.97e-2 &          862.613 & 3d$^6$ ($^5$D) 4s               & a $^6$D    &  3/2  \\
Fe II    &       2344.214 &        1.10e-1 &        1.14e-1 &                0 & 3d$^6$ ($^5$D) 4s               & a $^6$D    &  9/2  \\
Fe II*   &       2345.001 &        1.40e-1 &        1.53e-1 &          977.053 & 3d$^6$ ($^5$D) 4s               & a $^6$D    &  1/2  \\
Fe II*   &       2348.834 &        3.40e-2 &               &         1872.567 & 3p$^6$ 3d$^7$                   & a $^4$F    &  9/2  \\
Fe II*   &       2349.022 &        9.90e-2 &        8.98e-2 &          667.683 & 3d$^6$ ($^5$D) 4s               & a $^6$D    &  5/2  \\
Fe II*   &       2359.828 &               &        6.79e-2 &          862.613 & 3d$^6$ ($^5$D) 4s               & a $^6$D    &  3/2  \\
Fe II*   &       2360.721 &        2.00e-2 &               &         1872.567 & 3p$^6$ 3d$^7$                   & a $^4$F    &  9/2  \\
Fe II*   &       2365.552 &        5.10e-2 &        4.95e-2 &           384.79 & 3d$^6$ ($^5$D) 4s               & a $^6$D    &  7/2  \\
Fe II    &       2367.591 &               &        2.16e-5 &                0 & 3d$^6$ ($^5$D) 4s               & a $^6$D    &  9/2  \\
Fe II    &       2374.461 &        2.80e-2 &        3.13e-2 &                0 & 3d$^6$ ($^5$D) 4s               & a $^6$D    &  9/2  \\
Fe II    &       2382.765 &        3.90e-1 &        3.20e-1 &                0 & 3d$^6$ ($^5$D) 4s               & a $^6$D    &  9/2  \\
Fe II*   &       2389.358 &        8.06e-2 &        8.25e-2 &           384.79 & 3d$^6$ ($^5$D) 4s               & a $^6$D    &  7/2  \\
Co II*   &       2389.646 &        2.40e-1 &               &          3350.58 & 3p$^6$ 3d$^7$ (4F) 4s           & a $^5$F    &   5   \\
He I*    &       3188.655 &        2.57e-3 &               &      159855.9726 & 1s 2s                     & $^3$S      &   1   \\
He I*    &       3188.666 &        7.70e-3 &               &      159855.9726 & 1s 2s                     & $^3$S      &   1   \\
He I*    &       3188.667 &        1.28e-2 &               &      159855.9726 & 1s 2s                     & $^3$S      &   1   \\
Ti II    &       3242.918 &        1.83e-1 &        2.32e-1 &                0 & 3s$^2$ 3p$^6$ 3d$^2$ ($^3$F) 4s       & a $^4$F    &  3/2  \\
Na I     &       3303.319 &        9.00e-3 &        9.20e-3 &                0 & 3s                        & $^2$S      &  1/2  \\
Na I     &       3303.929 &        4.46e-3 &        4.60e-3 &                0 & 3s                        & $^2$S      &  1/2  \\
Ti II    &       3384.730 &        2.81e-1 &        3.58e-1 &                0 & 3s$^2$ 3p$^6$ 3d$^2$ ($^3$F) 4s       & a $^4$F    &  3/2  \\
Fe I     &       3720.993 &        4.11e-2 &        4.11e-2 &                0 & 3d$^6$ 4s$^2$                   & a $^5$D    &   4   \\
Fe I     &       3861.006 &        2.17e-2 &        2.17e-2 &                0 & 3d$^6$ 4s$^2$                   & a $^5$D    &   4   \\
He I*    &       3889.748 &        2.15e-2 &               &      159855.9726 & 1s 2s                     & $^3$S      &   1   \\
He I*    &       3889.751 &        3.58e-2 &               &      159855.9726 & 1s 2s                     & $^3$S      &   1   \\
Ca II    &       3934.775 &        6.82e-1 &        6.27e-1 &                0 & 4s                        & $^2$S      &  1/2  \\
Ca II    &       3969.590 &        3.30e-1 &        3.12e-1 &                0 & 4s                        & $^2$S      &  1/2  \\
CH       &       4301.523 &               &               &                  & X $^2\Pi$ (0)                  & R$_2$      &  1/2  \\
Na I     &       5891.583 &        6.41e-1 &        6.41e-1 &                0 & 3s                        & $^2$S      &  1/2  \\
Na I     &       5897.558 &        3.20e-1 &        3.20e-1 &                0 & 3s                        & $^2$S      &  1/2  \\
\end{supertabular}
}

\section{Column densities}

\begin{table}
\caption{Column densities towards HD 37061.}
\begin{tabular}{lrlrcc}
\hline
 element  & E [cm$^{-1}$]    & configuration           & term   &   J   & column density        [cm$^{-2}$]    \\
 \hline
As II    &              0 & 4s$^2$ 4p$^2$                 & $^3$P     &   0   &      6.72e11 $\pm$      2.0e11 \\
Ca II    &              0 & 4s                      & $^2$S     &  1/2  &      8.03e11 $\pm$      1.0e11 \\
Cd II    &              0 & 5s                      & $^2$S     &  1/2  &      2.21e11 $\pm$      7.5e09 \\
CH       &              0 & X $^2\Pi$ (0)           & R$_2$     &  1/2  &      3.09e12 $\pm$      1.3e12 \\
Cl I     &              0 & 3s$^2$ 3p$^5$                 & $^2$P$^o$ &  3/2  &      3.78e12 $\pm$      7.3e11 \\
Co II*   &        3350.58 & 3p$^6$ 3d$^7$ ($^4$F) 4s         & a $^5$F   &   5   &      2.37e11 $\pm$      5.6e10 \\
C I      &              0 & 2s$^2$ 2p$^2$                 & $^3$P     &   0   &      3.14e13 $\pm$      6.1e12 \\
C II     &              0 & 2s$^2$ ($^1$S) 2p             & $^2$P$^o$ &  1/2  &      8.87e17 $\pm$      2.0e17 \\
C I*     &         16.417 & 2s$^2$ 2p$^2$                 & $^3$P     &   1   &      5.09e13 $\pm$      1.2e13 \\
C I**    &         43.414 & 2s$^2$ 2p$^2$                 & $^3$P     &   2   &      4.24e13 $\pm$      7.2e12 \\
Fe I     &              0 & 3p$^6$ 3d$^6$ 4s$^2$             & a $^5$D   &   4   &      7.15e11 $\pm$      1.6e11 \\
Fe II    &              0 & 3d$^6$ ($^5$D) 4s             & a $^6$D   &  9/2  &      1.25e15 $\pm$      1.3e14 \\
Fe II*   &         384.79 & 3d$^6$ ($^5$D) 4s             & a $^6$D   &  7/2  &      1.69e12 $\pm$      1.4e11 \\
Fe II*   &        667.683 & 3d$^6$ ($^5$D) 4s             & a $^6$D   &  5/2  &      5.52e11 $\pm$      8.2e10 \\
Fe II*   &        862.613 & 3d$^6$ ($^5$D) 4s             & a $^6$D   &  3/2  &      4.39e11 $\pm$      1.1e11 \\
Fe II*   &        977.053 & 3d$^6$ ($^5$D) 4s             & a $^6$D   &  1/2  &      4.64e11 $\pm$      7.2e10 \\
Fe II*   &       1872.567 & 3p$^6$ 3d$^7$                 & a $^4$F   &  9/2  &      1.74e13 $\pm$      2.2e12 \\
Ge II    &              0 & 4s$^2$ 4p                  & $^2$P$^o$ &  1/2  &      2.03e12 $\pm$      1.5e11 \\
He I*    &    159855.9726 & 1s 2s                   & $^3$S     &   1   &      1.24e13 $\pm$      1.0e12 \\
Kr I     &              0 & 4s$^2$ 4p$^6$                 & $^1$S     &   0   &      5.66e12 $\pm$      9.6e11 \\
Mg II    &              0 & 2p$^6$ 3s                  & $^2$S     &  1/2  &      5.02e15 $\pm$      1.2e14 \\
Mn II    &              0 & 3s$^2$ 3p$^6$ 3d$^5$ ($^6$S) 4s     & $^7$S     &   3   &      7.15e13 $\pm$      4.8e13 \\
Na I     &              0 & 3s                      & $^2$S     &  1/2  &      4.01e12 $\pm$      2.1e12 \\
Ni II    &              0 & 3s$^2$ 3p$^6$ 3d$^9$             & $^2$D     &  5/2  &      6.54e13 $\pm$      7.8e12 \\
Ni II*   &         8393.9 & 3p$^6$ 3d$^8$ ($^3$F) 4s         & $^4$F     &  9/2  &      2.36e12 $\pm$      1.1e11 \\
N I*     &      19233.177 & 2s$^2$ 2p$^3$                 & $^2$D$^o$ &  3/2  &      4.76e12 $\pm$      1.6e12 \\
N I**    &      19224.464 & 2s$^2$ 2p$^3$                 & $^2$D$^o$ &  5/2  &      5.75e12 $\pm$      1.2e12 \\
O I      &              0 & 2s$^2$ 2p$^4$                 & $^3$P     &   2   &      1.46e18 $\pm$      5.9e16 \\
O I*     &        158.265 & 2s$^2$ 2p$^4$                 & $^3$P     &   1   &      3.77e14 $\pm$      2.3e13 \\
O I**    &        226.977 & 2s$^2$ 2p$^4$                 & $^3$P     &   0   &      2.61e14 $\pm$      6.6e12 \\
P II     &              0 & 3s$^2$ 3p$^2$                 & $^3$P     &   0   &      2.48e14 $\pm$      3.4e12 \\
P III    &              0 & 3s$^2$ ($^1$S) 3p             & $^2$P$^o$ &  1/2  &      1.23e13 $\pm$      1.3e12 \\
Si II    &              0 & 3s$^2$ ($^1$S) 3p             & $^2$P$^o$ &  1/2  &      1.34e16 $\pm$      2.2e15 \\
Si II*   &         287.24 & 3s$^2$ 3p                  & $^2$P$^o$ &  3/2  &      7.70e13 $\pm$      7.9e12 \\
S I      &              0 & 3s$^2$ 3p$^4$                 & $^3$P     &   2   &      2.07e12 $\pm$      4.8e11 \\
S II     &              0 & 3s$^2$ 3p$^3$                 & $^4$S$^o$ &  3/2  &      2.50e15 $\pm$      6.2e13 \\
S III*   &         298.69 & 3s$^2$ 3p$^2$                 & $^3$P     &   1   &      6.66e14 $\pm$      3.2e13 \\
S III*   &         833.08 & 3s$^2$ 3p$^2$                 & $^3$P     &   2   &      1.33e14 $\pm$      2.3e12 \\
Ti II    &              0 & 3s$^2$ 3p$^6$ 3d$^2$ ($^3$F) 4s     & a $^4$F   &  3/2  &      2.44e11 $\pm$      2.4e09 \\
\hline
\end{tabular}
\end{table}

\begin{figure}[tb]
    \centering
    \includegraphics[width=\textwidth,viewport=1 1 811 537,clip]{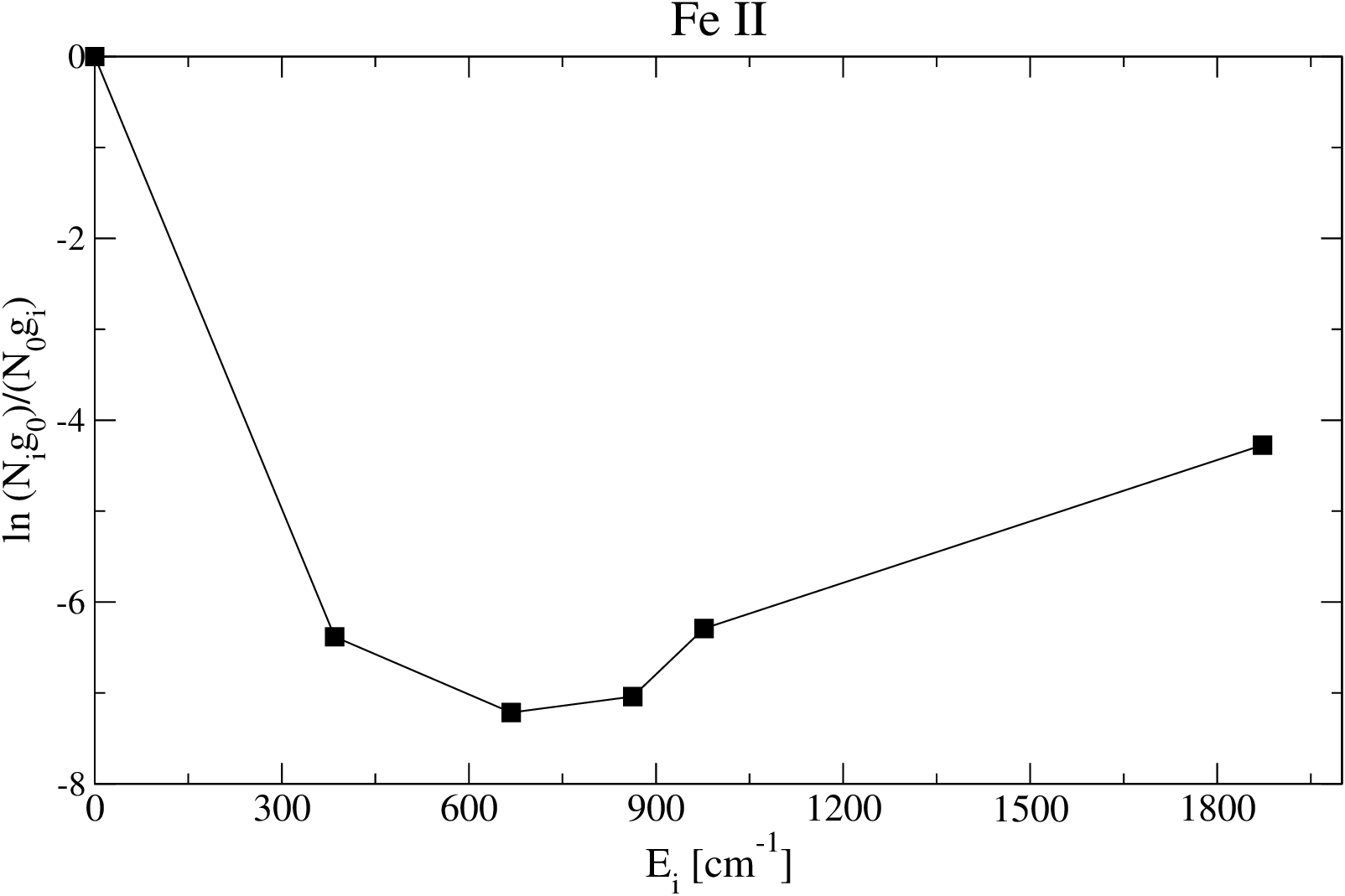}
    \caption{ Population of the excited Fe II levels. The points do not form a straight line, so the population of Fe II excited states does not follow the Boltzmann distribution.
    }
    \label{FeII}
\end{figure}

Column densities were calculated using the profile fitting technique. The Point Spread Function (PSF) for the STIS spectrograph was adopted from Quijano {\it et~al.} (2003).
The oscillator strength from Morton (2000 and 2003) was preferred over the NIST oscillator
strength in the calculation of column densities. Miller {\it et~al.} (2007) have observationally determined the new oscillator strength for the Fe II 2234 \AA\ line.

Table 2 presents the average column density for observed atoms and ions in the direction of HD 37061. Column densities were averaged over absorption lines of the same element and configuration. 

According to the Boltzmann distribution
\begin{equation}
\frac{N_i}{N_0} = \frac{g_i}{g_0}\exp{\left( - \frac{E_i-E_0}{kT} \right)} \;\;\; \Rightarrow \;\;\;  \ln{\frac{N_i g_0}{N_0 g_i}} = - \frac{E_i-E_0}{kT}
\end{equation}
the logarithm of column density on an excited level vs. energy should form a straight line.
Figure 2 shows that the occupation of excited levels of Fe II does not follow the Boltzmann distribution. We infer that fluorescence is responsible for the occupation of excited Fe II states.

The column density obtained from the Fe II line at 2367.59 \AA\ is 7 times larger than from other Fe II absorption lines. We suspect that the oscillator strength may be inaccurate or the line is a blend.

\section{Molecular hydrogen}

 \begin{figure}[tb]
    \centering
    \includegraphics[width=\textwidth,viewport=1 1 811 537,clip]{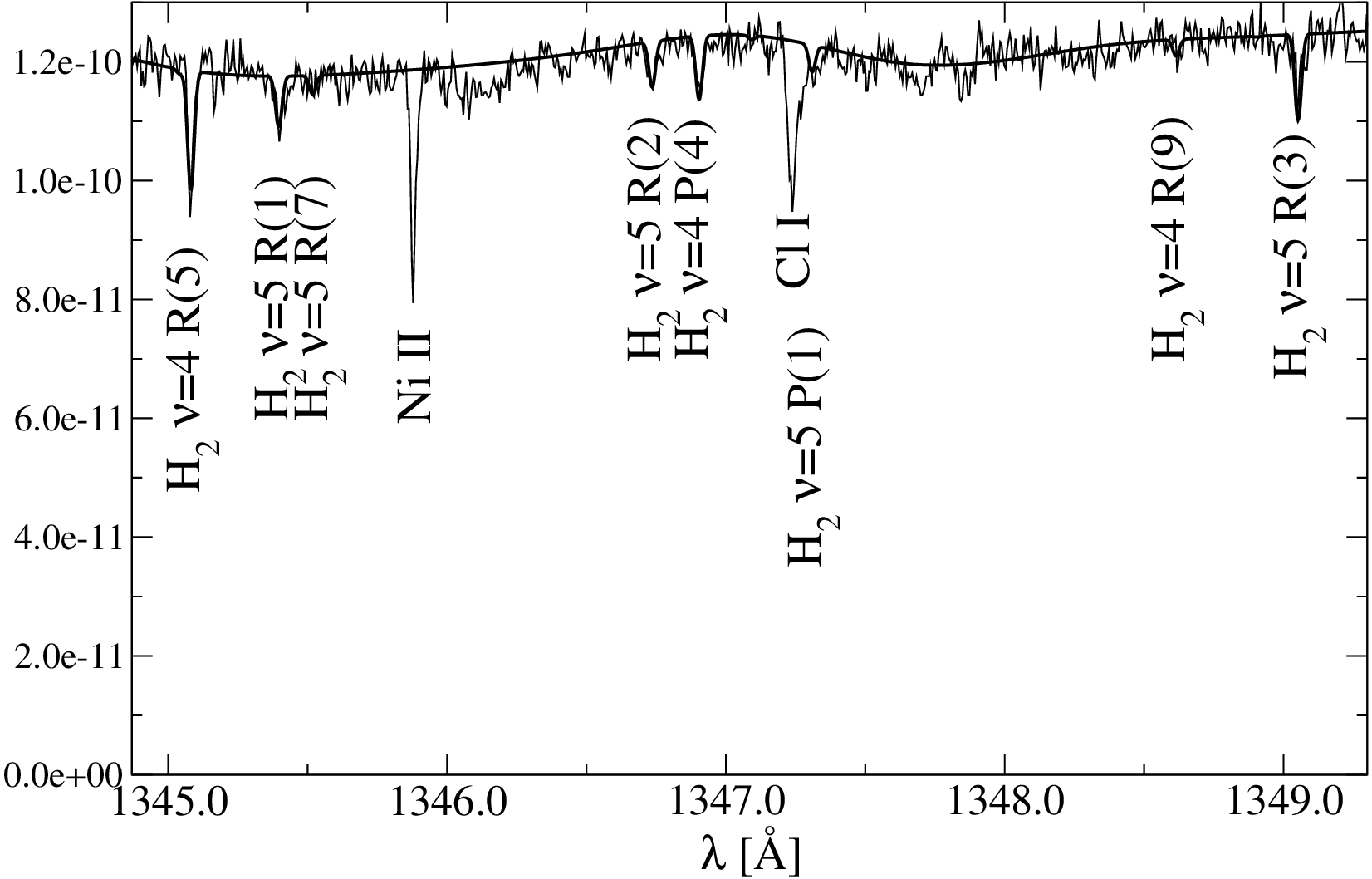}
    \caption{ HST spectrum towards HD 37061 with H$_2$ absorption lines from excited
     vibrational levels. Thick line shows the H$_2$ spectrum fitted to the averaged
     STIS spectrum.
    }
    \label{H2}
 \end{figure}
  
 \begin{table}
\caption{Observed H$_2$ column densities [cm$^{-2}$] of the ground electronic level (X) towards HD 37061. 
Column density standard deviation is shown in the line folowing the column density.
}
\begin{tabular}{l|rrrrrrrrrr}
\hline
J$\backslash\nu$ & 1 & 2 & 3 & 4 & 5 \\
\hline
0 & --- & 1.34e+12 & 1.03e+12 & 7.73e+11 & 1.75e+11 \\
  & --- & $\pm$5.7e+11 & $\pm$3.4e+11 & $\pm$6.5e+11 & $\pm$1.2e+11 \\
1 & --- & 1.58e+13 & 7.90e+12 & 5.37e+12 & 4.53e+12 \\
  & --- & $\pm$1.3e+12 & $\pm$4.7e+11 & $\pm$2.8e+11 & $\pm$1.3e+12 \\
2 & 3.53e+12 & 9.18e+12 & 4.12e+12 & 2.84e+12 & 3.12e+12 \\
& $\pm$3.0e+12 & $\pm$1.7e+12 & $\pm$1.1e+12 & $\pm$5.0e+11 & $\pm$4.8e+11 \\
3 & 3.56e+13 & 1.99e+13 & 9.86e+12 & 7.36e+12 & 5.90e+12 \\
& $\pm$1.7e+13 & $\pm$1.7e+12 & $\pm$2.6e+12 & $\pm$5.2e+11 & $\pm$6.8e+11 \\
4 & 4.48e+12 & 6.78e+12 & 4.16e+12 & 2.87e+12 & 3.26e+12 \\
& $\pm$6.0e+12 & $\pm$1.1e+12 & $\pm$2.1e+11 & $\pm$2.9e+11 & $\pm$9.0e+11 \\
5 & 9.67e+12 & 1.53e+13 & 7.35e+12 & 4.77e+12 & 6.19e+12 \\
& $\pm$8.0e+12 & $\pm$2.7e+12 & $\pm$1.8e+11 & $\pm$1.4e+11 & $\pm$6.4e+11 \\
6 & 1.13e+13 & 4.66e+12 & 2.13e+12 & 1.49e+12 & 5.34e+11 \\
& $\pm$6.6e+11 & $\pm$1.8e+12 & $\pm$7.9e+11 & $\pm$5.7e+11 & $\pm$7.0e+11 \\
7 & 8.29e+12 & 1.07e+13 & 4.47e+12 & 3.97e+12 & 3.70e+12 \\
& $\pm$6.6e+12 & $\pm$2.2e+12 & $\pm$5.1e+11 & $\pm$2.2e+11 & $\pm$9.5e+11 \\
8 & 8.67e+12 & 2.34e+12 & 1.85e+12 & 1.19e+12 & 1.52e+12 \\
& $\pm$8.7e+11 & $\pm$8.6e+10 & $\pm$9.8e+09 & $\pm$2.7e+11 & $\pm$5.4e+11 \\
9 & 1.24e+13 & 6.46e+12 & 3.85e+12 & 2.35e+12 & 1.31e+12 \\
& $\pm$1.9e+12 & $\pm$5.6e+11 & $\pm$2.6e+11 & $\pm$1.0e+12 & $\pm$3.0e+11 \\
\hline
\end{tabular}
\end{table}

 A rich spectrum of vibrationally excited H$_2$ was described by 
Meyer {\it et~al.} (2001) in the direction of HD 37903. They have also noticed, that
vibrationally excited H$_2$ is also present in the direction of HD 37061.
An example spectrum of vibrationally excited molecular hydrogen is presented on Fig. 3.
 
  \begin{figure}[tb]
    \centering
    \includegraphics[width=\textwidth,viewport=1 1 811 537,clip]{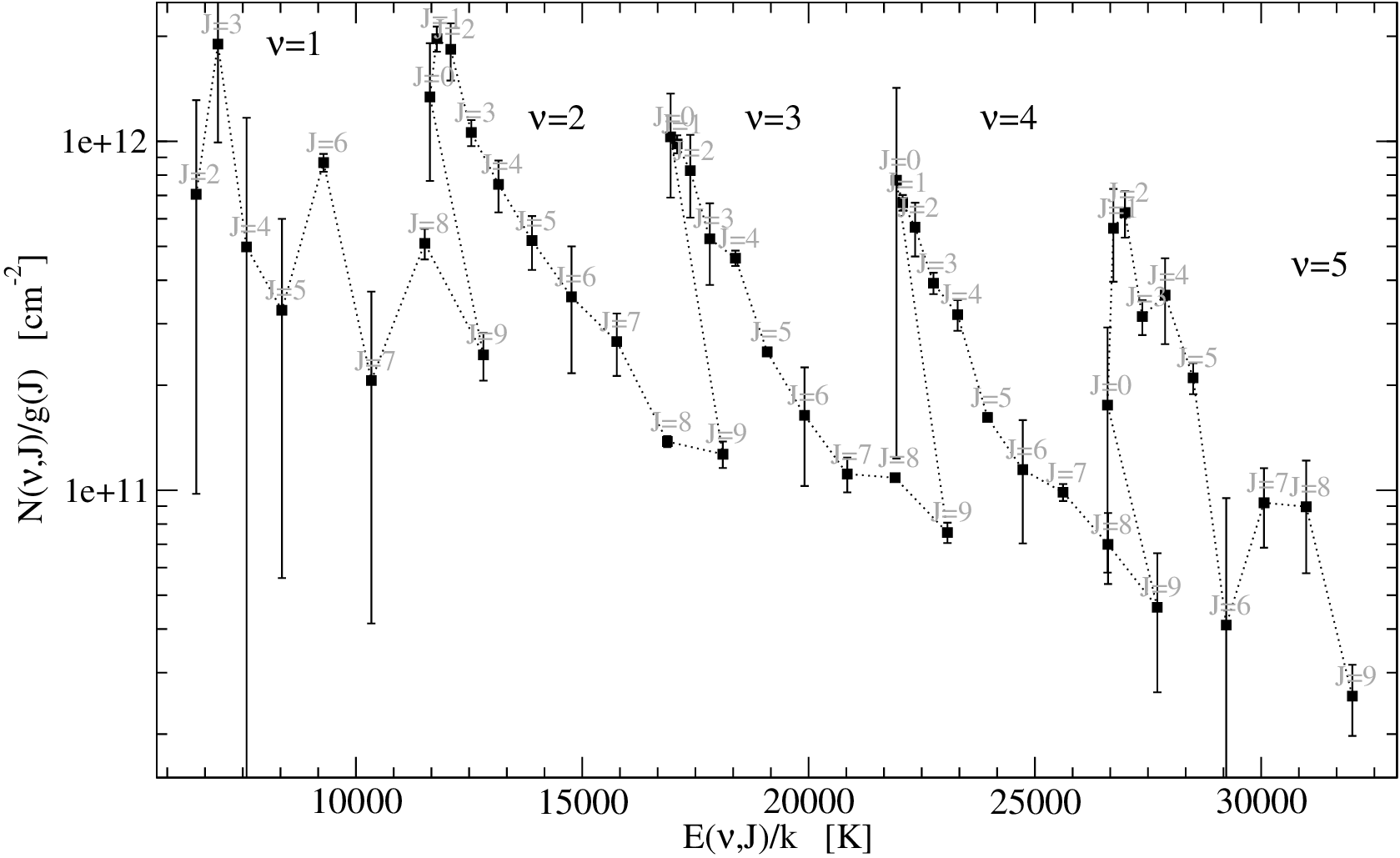}
    \caption{ Occupation of H$_2$ X ro-vibrational levels towards HD 37061.
    }
    \label{H2levels}
 \end{figure}
 
 We have used the H$_2$ spectral line database available at VO-Paris -- {\it "Basecol molecules service"}. The database is based on Abgrall {\it et~al.} (1994) and lists 33764 H$_2$ lines. We have selected 690 H$_2$ lines for $\nu$=1--5 and J=0--9 located in the STIS spectrum range 1163 \AA \ -- 1356.6 \AA. All these lines were simultaneously fitted to the observed spectrum to obtain
column densities of vibrationally--rotational H$_2$ levels.
For some H$_2$ levels with shallow lines the standard deviation of column density (see Table 3) is comparable to the column density value. 
Strong lines have column density error of $\sim$10\%.

The observed ortho/para H$_2$ ratio for $\nu$=2, 3 and 4 levels (levels with the best accuracy of column densities) is equal to 2.7. This value is close to the laboratory (at standard
temperature and pressure) ortho/para H$_2$ ratio 3:1. Our ortho/para H$_2$ ratio is higher than ortho/para H$_2$ equal to 1.45 observed by Meyer {\it et~al.}(2001) in the direction of HD 37903.
 
The rotational temperature of the H$_2$ molecule was calculated for all vibrational levels. Applying the linear regression method to the Boltzmann distribution in the form
\begin{equation}
 \ln{\frac{N_i}{g_i}} = \ln{\frac{N_0}{g_0}} - \frac{E_i-E_0}{kT}
\end{equation}
 we infer that rotation temperature $T_{rot}=3001 \pm 1232 $ K for the $\nu$=1 vibrational level, $T_{rot}=2280 \pm 166 $ K ($\nu$=2), $T_{rot}=2225 \pm 194 $ K ($\nu$=3), 
 $T_{rot}=2042 \pm 151 $ K ($\nu$=4), $T_{rot}=2068 \pm 463 $ K ($\nu$=5).

The target star HD 37061 was not observed by FUSE (Rachford {\it et~al.}, 2009), so the total H$_2$ column density on all levels remains unknown.

\section{Conclusions}
  
  The physical conditions in the M 43 emission nebula lead to observable amount of ions in excited levels with forbidden transitions to ground energy level. 
  The rich spectrum of vibrationally--rotational H$_2$ lines allows to determine the 
observed ortho/para H$_2$ ratio (2.7) and the rotational temperature of vibrational levels 1 -- 5 in the direction of HD 37061.

\section*{Acknowledgments}
  I would like to thank Daniel Welty for pointing my attention to the H$_2$ lines.
  This publication is based on observations made with the NASA/ESA Hubble Space Telescope, obtained from the data archive at the Space Telescope Institute. STScI is operated by the association of Universities for Research in Astronomy, Inc. under the NASA contract NAS 5-26555. We also used observations made with ESO VLT/UVES at the Paranal Observatory.
  The research was supported by University of Gda\'nsk grant BW/5400-5-0167-9.

\end{document}